\documentclass[twocolumn,twocolappendix]{aastex631}

\usepackage{amsmath}
\usepackage{newtxtext}
\usepackage{newtxmath}
\usepackage{graphicx}
\usepackage{subfigure}
\usepackage{booktabs}

\newcommand{\gcc}{\,g\,cm$^{-3}$}	
\newcommand{\ergs}{\,erg\,s$^{-1}$}	
\newcommand{\Ms}{\,M$_\odot$}       
\newcommand{\GK}{\,GK}	            

\begin{document}

\title{On the Nucleosynthesis in Accretion-Induced Collapse of White Dwarfs}
\shorttitle{On the Nucleosynthesis in Accretion-Induced Collapse of White Dwarfs}
\submitjournal{ApJ}
\shortauthors{Yip et al.}

\author[0000-0002-3311-5387]{Chun-Ming Yip}
\affiliation{Department of Physics and Institute of Theoretical Physics, The Chinese University of Hong Kong, Shatin, N.T., Hong Kong S.A.R., People's Republic of China}

\author[0000-0002-1971-0403]{Ming-Chung Chu}
\affiliation{Department of Physics and Institute of Theoretical Physics, The Chinese University of Hong Kong, Shatin, N.T., Hong Kong S.A.R., People's Republic of China}

\author[0000-0002-4972-3803]{Shing-Chi Leung}
\affiliation{Department of Mathematics and Physics, SUNY Polytechnic Institute, 100 Seymour Road, Utica, NY 13502, USA}

\author[0000-0002-4638-5044]{Lap-Ming Lin}
\affiliation{Department of Physics and Institute of Theoretical Physics, The Chinese University of Hong Kong, Shatin, N.T., Hong Kong S.A.R., People's Republic of China}

\begin{abstract}

It has long been hypothesized that accretion-induced collapse (AIC) of white dwarfs contribute to heavy chemical elements production in the universe. We present one-dimensional neutrino-radiative hydrodynamic simulations of AIC followed by post-processing nucleosynthesis calculations of the ejecta. A proto-neutron star is formed after the AIC, and a neutrino burst with peak luminosity $\sim10^{53}${\ergs}, comparable to that of a core-collapse supernova (CCSN), is emitted. The ejecta mass of AIC could be up to $\sim10^{-2}${\Ms}, and the first neutron-capture peak elements (Sr, Y, and Zr) could be abundantly synthesized, with an overproduction of $\sim10^{6}$ relative to the solar abundances. The yield of $^{56}\text{Ni}$ could be up to at most $\sim10^{-3}${\Ms}, suggesting that the electromagnetic light curve associated with AIC is at least 2 orders dimmer than those associated with Type Ia supernovae (Type Ia SN). The inferred upper bound of AIC event rate, from nucleosynthesis calculations, is at most $\sim10\,\%$ relative to those of CCSNe and Type Ia SNe.

\end{abstract}


\section{Introduction}

\subsection{Accretion-induced Collapse of White Dwarfs}

A typical carbon-oxygen white dwarf (WD) explodes as a Type Ia supernova (Type Ia SN) when it approaches the Chandrasekhar limit through mass accretion. \cite{1991ApJ...367L..19N}, on the other hand, proposed that electron capture could be triggered at the core of an oxygen-neon-magnesium WD and reduce the electron degeneracy pressure above certain threshold densities. In this scenario, the WD would undergo gravitational collapse, thus called accretion-induced collapse (AIC), to form a proto-neutron star (PNS) rather than a thermonuclear explosion that results in a Type Ia SN. Several studies revisited the AIC with oxygen-neon WDs being the progenitors \citep{2018RAA....18...36W,2018MNRAS.481..439W,2019MNRAS.484..698R}. A recent work by \cite{2023arXiv230617381M} also discusses the initial conditions of super-Chandrasekhar AIC progenitors. A similar electron capture mechanism was also employed to account for the gravitational collapse and explosion for super-asymptotic giant branch (AGB) stars of $8-10${\Ms} as electron-capture supernovae (ECSNe) \citep{2019PASA...36....6L,2020ApJ...889...34L}.

Neutron stars follow a bimodal mass distribution (e.g., \citep{2010ApJ...719..722S,2019ApJ...876...18F}). AIC is a possible formation channel of the low-mass neutron stars since the AIC progenitor mass does not exceed the Chandrasekhar limit. Moreover, it was proposed that AIC could be an explanation of multi-wavelength electromagnetic (EM) transients, including fast radio burst \citep{2013ApJ...762L..17P,2016ApJ...830L..38M,2019ApJ...886..110M}, X-ray transient \citep{2019ApJ...877L..21Y}, and gamma-ray burst \citep{1998ApJ...494L.163Y}. Nevertheless, these transients are known to be faint and short-lived due to the low yield of $^{56}\text{Ni}$ up to at most $\sim10^{-2}${\Ms} \citep{2009MNRAS.396.1659M}. Although there is no confirmed EM observation of AIC yet, such an event may be alternatively identified through multi-messenger observations. The axisymmetric general relativistic simulations of AIC by \cite{2010PhRvD..81d4012A} predicted that the gravitational-wave signal of a galactic AIC event would be detectable. Multidimensional simulations of AIC with neutrino transport were also performed by \cite{2006ApJ...644.1063D,2007ApJ...669..585D} to study the effects of neutrino-matter interactions on explosion dynamics. A recent work on three-dimensional general relativistic simulations with neutrino transport by \cite{2023MNRAS.tmp.2326M} reported that the neutrino burst associated with AIC is as bright as those emitted by core-collapse supernovae (CCSNe). The next generation of neutrino detectors, such as \texttt{JUNO} \citep{2022NIMPA104267435Y} and \texttt{DUNE} \citep{2022NIMPA104167217F}, may help identify AICs. The upcoming multi-messenger observations offer excellent opportunities to study AIC with unprecedented details.

\subsection{Supernova Nucleosynthesis} \label{subsection: Supernova Nucleosynthesis}

It is widely believed that about half of the heavy chemical elements in nature are produced by astrophysical rapid neutron-capture process ($r$-process) nucleosynthesis (see the reviews by e.g., \cite{2019PrPNP.107..109K,2021RvMP...93a5002C}). AIC was hypothesized to be a promising $r$-process nucleosynthesis site for its neutron-rich outflow \citep{1992ApJ...391..228W,1999ApJ...516..892F}. The event rate of AICs, accordingly, could be inferred from its production curves. The study by \cite{2010MNRAS.409..846D} determined that the first neutron-capture peak elements (Sr, Y, and Zr) could be synthesized through AIC, and the event rate of AICs is predicted to be $\sim10^{-2}$ of the event rate of Type Ia SNe. Without taking AIC into account, the review by \cite{2020ApJ...900..179K} argued that the first neutron-capture peak elements are sufficiently produced by ECSNe and AGB stars. The calculations by \cite{2009ApJ...695..208W,2011ApJ...726L..15W} deduced the overproduction of the first neutron-capture peak elements through an ECSN event, and hence imposed a constraint on the event rate of ECSNe relative to other CCSNe.

In this paper, we perform one-dimensional hydrodynamic simulations of AIC with neutrino transport to investigate the properties of its mass outflow, as well as post-processing nucleosynthesis calculations to study its chemical elements production. In addition, we investigate the equation of state (EoS) dependence of ejecta properties and characteristic isotopes production of AIC. In the following discussion, we show that AIC may significantly contribute to the production of the first neutron-capture peak elements in nature based on the production curve. Based on our results, we derive a new constraint on the event rate of AICs relative to Type Ia SNe and CCSNe from our nucleosynthesis calculations. Due to the degeneracy of production patterns between AICs and ECSNe, the former constraints on the event rate of ECSNe suggested by \cite{2009ApJ...695..208W,2011ApJ...726L..15W} may also be affected after including the contribution from AICs.

This paper is organized as follows. In Section~\ref{section 2}, we introduce the computational setup of the neutrino radiation hydrodynamic simulations and nucleosynthesis calculations used. The numerical results of the simulations and calculations are presented in Section~\ref{section 3}. We summarize and discuss the implications of these results in Section~\ref{section 4} followed by our conclusions in Section~\ref{section 5}.

\section{Methodology} \label{section 2}

\subsection{Initial Models}

\begin{figure}[t!]
    \includegraphics[width=\columnwidth]{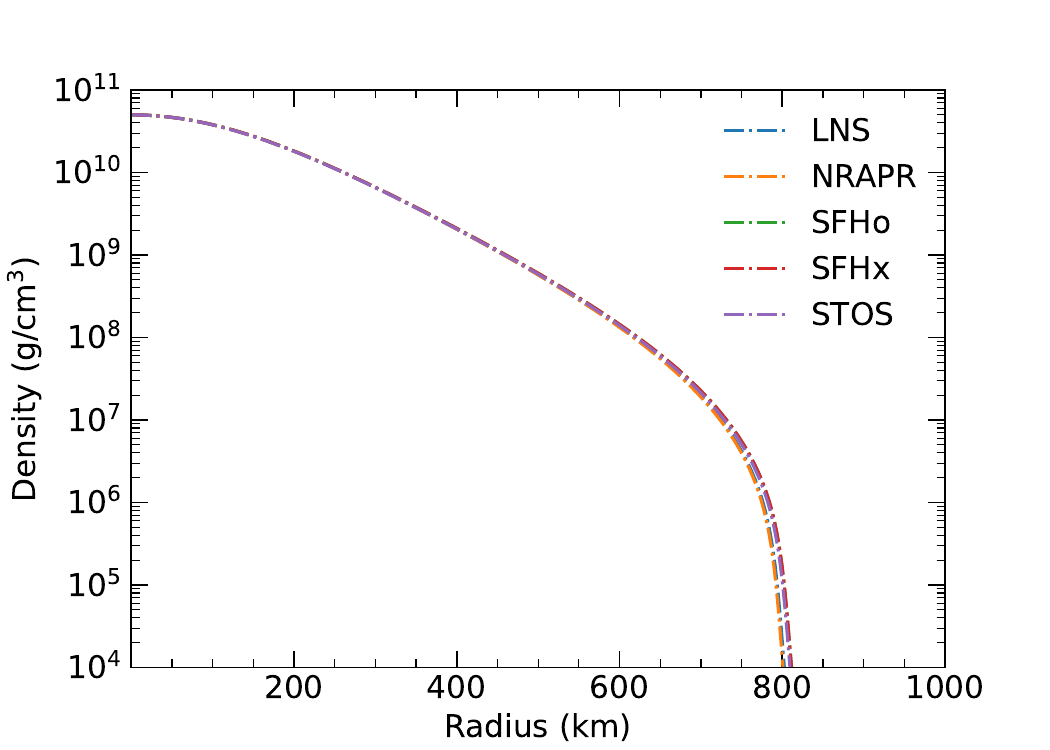}
    \caption{Density vs. radius of initial WD models using different EoSs.}
    \label{fig:initial density profile}
\end{figure}

We construct spherically symmetric WD models as AIC progenitors at Newtonian hydrostatic equilibrium. An initial parameterized temperature profile is assigned to the WD models following \cite{2006ApJ...644.1063D}:
\begin{equation}
    T(\rho) = T_c \left( \frac{\rho}{\rho_c} \right)^{0.35},
\end{equation}
where $T_c = 10${\GK} and $\rho_c = 5\times10^{10}${\gcc} are the central temperature and density of the WDs, respectively. We employ a grid setup to define the spatial resolution $r_j$ of the $j-$th grid, similar to that proposed by \cite{2016ApJ...831...81S}:
\begin{equation}
   r_j = A_f x_t \sinh{\left( \frac{j}{x_t} \right)}.
\end{equation}
We set $A_f=300${\,m} to be the finest resolution at the centre and $x_t=500$, so that the progenitors are contained by about $1,200$ grids. The density profile of the initial models is shown in Figure~\ref{fig:initial density profile}. With the given $T_c$ and $\rho_c$, these WDs have masses $\approx1.4${\Ms} and radii $\approx800${\,km} that are insensitive to the EoS choice (see Section~\ref{subsection: Equation of state} for more details).

\subsection{Hydrodynamics} \label{subsection: Hydrodynamics}

In this paper, the hydrodynamic simulations of AIC are performed using the one-dimensional Newtonian hydrodynamics code developed by \cite{2015MNRAS.454.1238L}. To mimic the general relativistic effects on the PNS with a very high compactness, we implement an effective relativistic potential and the lapse function to the Newtonian hydrodynamics following the Case A formalism in \cite{2006A&A...445..273M} \citep[see also][for the reference]{2020PhRvL.125e1102Z}.

For the electron capture that is responsible for the gravitational collapse of the progenitors, we follow the parameterized fitting formula suggested by \cite{2005ApJ...633.1042L} to determine the electron fraction $Y_e$ as a function of density $\rho$ before core bounce:
\begin{equation}
    \begin{aligned}
    Y_e(x) = & \frac{1}{2} \left( Y_1 + Y_2 \right) + \frac{x}{2} \left( Y_1 - Y_2 \right) + \\ & Y_3 \left[ 1 - |x| + \frac{4 |x| \left( |x| - \frac{1}{2} \right)}{|x| - 1} \right],
    \end{aligned}
\end{equation}
\begin{equation}
    x(\rho) = \max \left[ -1, \min \left( 1, \frac{2 \log{\rho} - \log{\rho_2} - \log{\rho_1}}{\log{\rho_2} - \log{\rho_1}} \right) \right].
\end{equation}
The fitting parameters $\rho_1$, $\rho_2$, $Y_1$, $Y_2$, and $Y_3$ are obtained from more sophisticated calculations, and they are used in several earlier simulations of AIC \citep{2019ApJ...884....9L,2019ApJ...883...13Z,2023ApJ...945..133C}. This parameterized fitting formula for electron capture is applied until the core bounce condition is satisfied when the entropy of the core exceeds $3 k_B$, where $k_B$ is the Boltzmann constant. Neutrino pressure is included in the neutrino trapping regime at $\rho \ge 2\times10^{12}${\gcc}. After the core bounce, we activate a neutron transport scheme (see Section~\ref{subsection: Neutrino transport}) that couples with the hydrodynamics.

In each model, we execute the hydrodynamic simulations up to $1${\,s} after the core bounce in order to determine the mass outflow properties of AIC.

\subsection{Equation of state} \label{subsection: Equation of state}

In an AIC simulation, a PNS with maximum temperature $> 10${\,MeV} and density $> 10^{14}${\gcc} forms through the gravitational collapse of the WD. Meanwhile, the electron fraction of the PNS could drop below $0.1$ as a consequence of neutronization. The properties of hot and dense nuclear matter with large isospin asymmetry are highly uncertain. The EoSs constructed for supernova simulations rely on theoretical predictions of nuclear matter properties under these extreme conditions that cannot be achieved experimentally. We adopt a few open-source EoS tables\footnote{\url{https://stellarcollapse.org/}} to investigate the effects of varying EoS on our numerical calculations. We employ the LNS and NRAPR EoSs \citep{PhysRevC.96.065802,PhysRevC.100.025803,2006PhRvC..73a4313C,STEINER2005325,LATTIMER1991331} as representative low stiffness models. For intermediate stiffness models, we choose the SFHo and SFHx EoS \citep{2010NuPhA.837..210H,2013ApJ...774...17S,moller1997nuclear,2012ApJ...748...70H}. The STOS EoS \citep{1998NuPhA.637..435S,2011ApJS..197...20S} is additionally used as a high stiffness model. The incompressibility $K$ and TOV limit $M_{\text{TOV}}$ of these models are listed in Table~\ref{tab:model stiffness and remarkable results}.

\subsection{Neutrino transport} \label{subsection: Neutrino transport}

Neutrinos are produced, absorbed, and scattered by matter after a PNS forms. The heating and cooling of the star due to neutrino-matter interactions are coupled with the hydrodynamics and influence the evolution of the star. During the post-bounce phase, we evaluate the neutrino transport using the isotropic diffusion source approximation (IDSA) \citep{2009ApJ...698.1174L,doi:10.1137/12089243X}.

The charged-current interactions between the electron-type neutrinos, namely $\nu_e$ and $\bar{\nu}_e$, and nucleons (neutron $n$ and proton $p$) are included as the major electron-type neutrinos production and absorption channels:
\begin{equation}
    p + e^- \longleftrightarrow \nu_e + n,
\end{equation}
\begin{equation}
    n + e^+ \longleftrightarrow \bar{\nu}_e + p.
\end{equation}
Besides, the elastic scattering channels between electron-type neutrinos and nucleons, alpha particles, and nuclei are included. To include the neutrino cooling effect through heavy flavor neutrinos $\nu_X$ emissions, we adopt the leakage scheme \citep{2003MNRAS.342..673R} for muon- and tau-type neutrinos together with a fitting formula for their pair-production rates \citep{1996ApJS..102..411I}.

\subsection{Nucleosynthesis}

In order to investigate the chemical elements production of AIC ejecta, we use the open-source code \texttt{torch} for nuclear reaction network calculation developed by \citet{2000ApJS..129..377T}. The network consists of about $1,200$ isotopes from neutron and proton to Sn (atomic number $Z = 50$). The nuclear masses, partition functions, and thermonuclear reaction rates, including $(n,\gamma), (n,p), (p,\gamma), (\alpha,n), (\alpha,p), (\alpha,\gamma)$, and their inverse processes, are obtained from \cite{2010ApJS..189..240C}. The weak decay rates of isotopes computed by \cite{2003PhRvC..67e5802M} are used\footnote{The half-life of $^{56}$Ni is set to be $7.605\times10^{5}$\,s instead \citep{1982ApJ...253..785A}.}. Some special nuclear reactions, such as triple-alpha process, are also attached to the network.

We perform the post-processing nucleosynthesis calculations by mapping the hydrodynamic and thermodynamic quantities of the ejected material, obtained from the hydrodynamic simulations of AIC done in Section~\ref{subsection: Hydrodynamics}, to about $30$ trajectories. After that, we evaluate the nucleosynthesis of these trajectories by solving the nuclear reaction network explicitly. We start the nucleosynthesis calculations when the temperature of the ejecta drops below $10${\GK}, at which the neutrino-matter interactions freeze. The initial mass fractions of isotopes for the ejected trajectories are assigned assuming nuclear statistical equilibrium (NSE). The mass fractions are then evolved by solving the network following the density and temperature evolution of each trajectory individually. Beyond the hydrodynamic simulation time, we extrapolate the density and temperature of the ejecta by $\rho(t) \propto t^{-3}$ and $T(t) \propto t^{-1}$ as functions of time $t$ to feature the free expansion of the ejected material. We update the network for $1${\,Gyr} to determine the stable chemical elements produced.

\section{Results} \label{section 3}

\begin{deluxetable*}{c|cc|cccccc}
    \label{tab:model stiffness and remarkable results}
    \caption{EoS properties and simulation results.}
    \tablehead{model & $K$ & $M_{\text{TOV}}$ & $M_{\text{ejecta}}$ & $M(^{56}\text{Ni})$ & $M(^{90}\text{Zr})$ & $f_{\text{CCSN}}$ & 
    $f_{\text{SNIa}}$ & }
    \startdata
         LNS   & 211   & 1.72 & 1.02e-03 & 1.16e-04 & 2.62e-05 & 1.09e-01 & 2.72e-01 \\
         NRAPR & 226   & 1.94 & 9.10e-04 & 4.02e-05 & 2.57e-05 & 1.11e-01 & 2.77e-01 \\
         SFHo  & 245.4 & 2.06 & 2.99e-02 & 3.56e-03 & 1.58e-04 & 1.81e-02 & 4.49e-02 \\
         SFHx  & 238.8 & 2.13 & 2.35e-02 & 4.86e-03 & 2.15e-04 & 1.33e-02 & 3.30e-02 \\
         STOS  & 281   & 2.23 & 2.60e-02 & 4.82e-03 & 6.41e-05 & 4.46e-02 & 1.11e-01 &  \\
    \enddata
    \tablecomments{Incompressibility $K$ (MeV) and TOV limit $M_{\text{TOV}}$ (M$_\odot$), the maximum mass of neutron stars, for different EoS models. The total ejecta mass $M_{\text{ejecta}}$, yield of $^{56}\text{Fe}$, and yield of $^{90}\text{Zr}$ for these models are shown in units of {\Ms}. The event rate of AIC relative to those of CCSNe $f_{\text{CCSN}}$ and Type Ia SNe $f_{\text{SNIa}}$ are evaluated using Equation~(\ref{eq:AIC rate to CCSN}) and Equation~(\ref{eq:AIC rate to Type Ia SN}), respectively.}
\end{deluxetable*}

\subsection{Formation of proto-neutron stars}

\begin{figure}[t!]
    \includegraphics[width=\columnwidth]{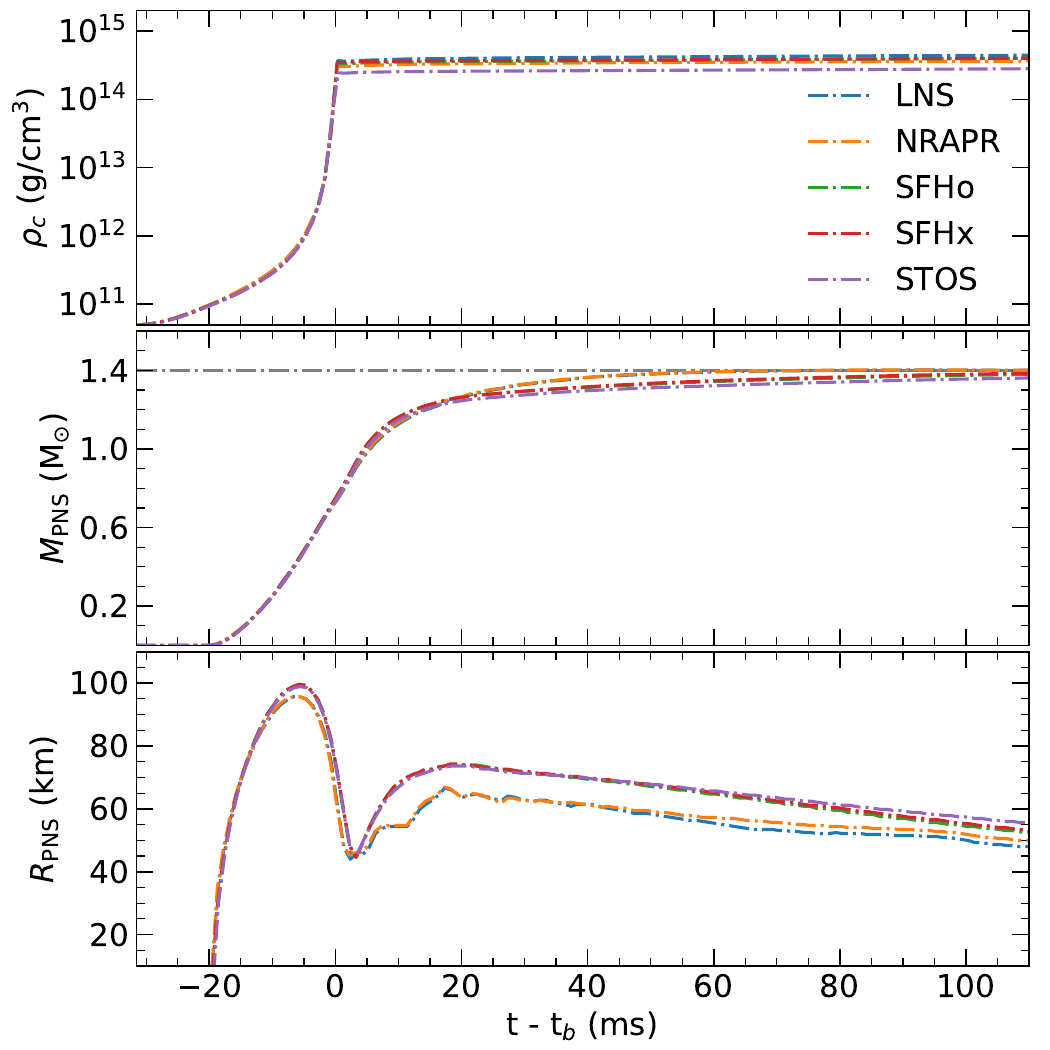}
    \caption{Time evolution of the central density $\rho_c$ (upper panel), PNS mass $M_{\text{PNS}}$ (middle panel), and PNS radius $R_{\text{PNS}}$ (lower panel) for all AIC models. After the core bounce, $M_{\text{PNS}}$ grows and approaches $1.4${\Ms} shown by the grey dash dotted horizontal line in the middle panel.}
    \label{fig:PNS evolution}
\end{figure}

After the simulation starts, the inner region of the WD starts to collapse owing to electron capture. The core bounce (see Section~\ref{subsection: Hydrodynamics}) occurs at $t_b\approx32${\,ms} after the start of simulation, indicating the formation of a PNS. The accretion shock induced by the core bounce propagates outward and breaks out from the WD surface at $t-t_b\approx100-200${\,ms} (EoS dependent) after the bounce. The time evolution of the central density, mass, and radius of the PNS, defined to be the inner region of the star with density $\rho \ge 10^{11}${\gcc}, are shown in Figure~\ref{fig:PNS evolution} for the SFHo model. After the core bounce, the PNS accretes surrounding matter and contracts in size. The accretion by the PNS is insensitive to the EoS, and the resulting PNS through AIC always grows to $M_{\text{PNS}}\approx1.4${\Ms} for all EoSs used in this work.

\subsection{Neutrino signals}

\begin{figure}[t!]
    \includegraphics[width=\columnwidth]{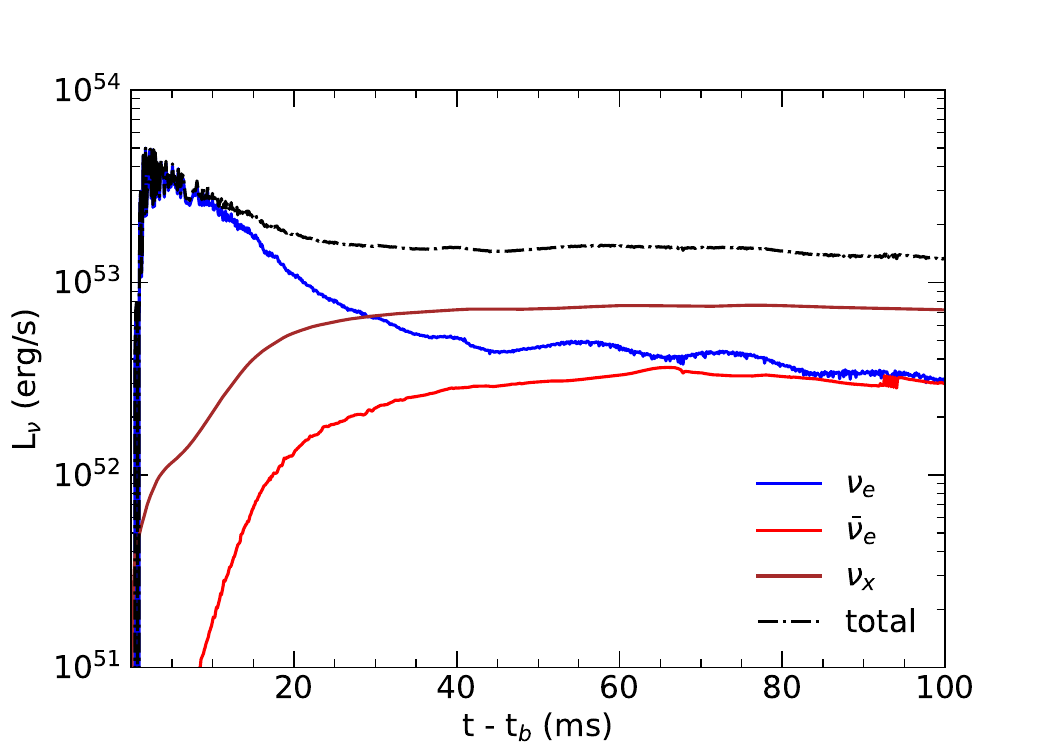}
    \caption{Time evolution of neutrino luminosity $L_\nu$ for the AIC model with the STOS EoS. The total luminosity combining all flavors is indicated by the black dashed-dotted curve.}
    \label{fig:neutrino luminosity}
\end{figure}

\begin{figure}[t!]
    \includegraphics[width=\columnwidth]{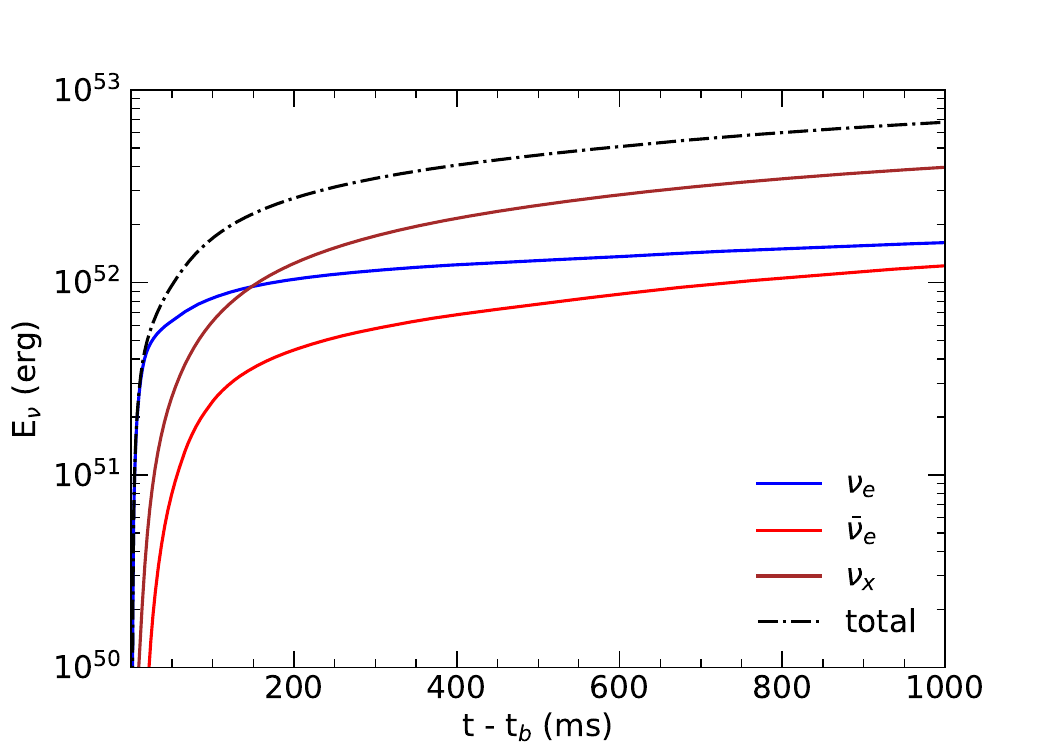}
    \caption{Cumulative energy loss through neutrino emission $E_\nu$ for the AIC model with the STOS EoS. The total energy loss combining all flavors is indicated by the black dashed-dotted curve.}
    \label{fig:neutrino energy loss}
\end{figure}

After the formation of a PNS, a neutronization burst of electron neutrinos is emitted. The time evolution of neutrino signals of AIC after the core bounce is shown in Figure~\ref{fig:neutrino luminosity} for the STOS model. The peak luminosity of the neutronization burst is $\approx5\times10^{53}${\ergs}. The PNS emits neutrino fluxes in all flavours subsequently, and the luminosity of each flavour remains at $\sim10^{52}${\ergs} until the end of the simulation. The cumulative energy released through neutrino emission for the same model is shown in Figure~\ref{fig:neutrino energy loss}. The neutrino signals for other models are found to be quite similar to those displayed in Figure~\ref{fig:neutrino luminosity} and Figure~\ref{fig:neutrino energy loss}. At the end of the simulations, the total energy released through neutrino emission is $\approx6 (7)\times10^{52}${\,erg} for the models using low (intermediate and high) stiffness EoSs.

\subsection{Nucleosynthesis}  \label{subsection: Nucleosynthesis}

\begin{figure}[t!]
    \includegraphics[width=\columnwidth]{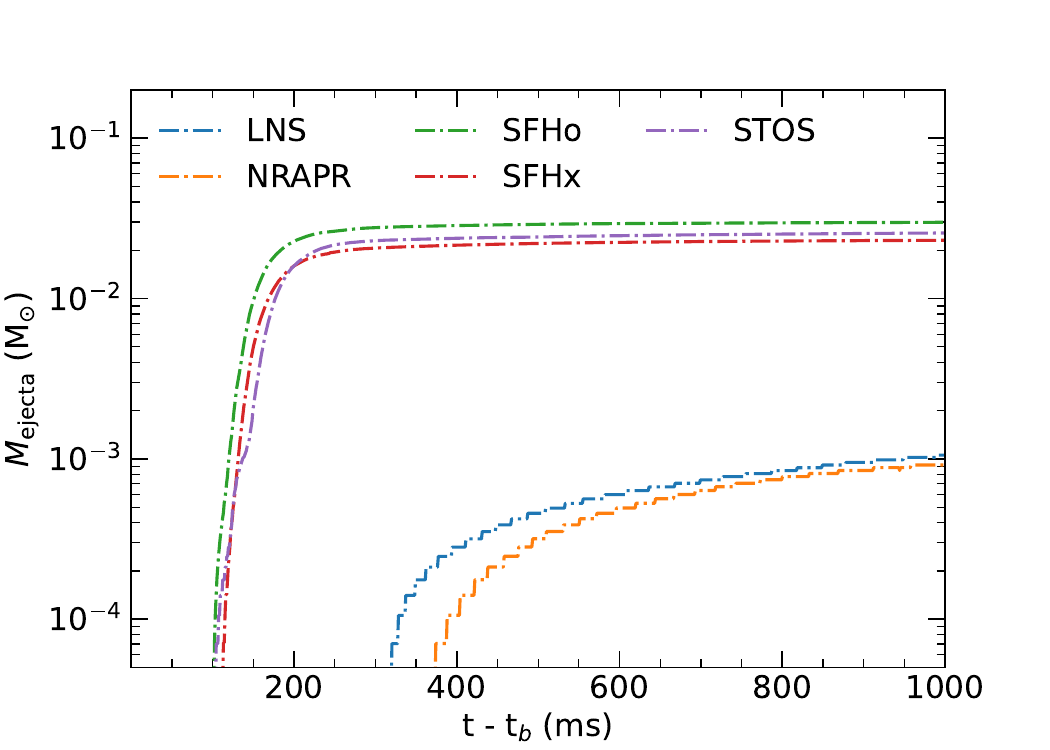}
    \caption{Time evolution of the total ejecta mass for all models.}
    \label{fig:ejecta mass evolution}
\end{figure}

\begin{figure*}[t!]
    \centering
    \subfigure[LNS model]  {\includegraphics[width=0.45\textwidth]{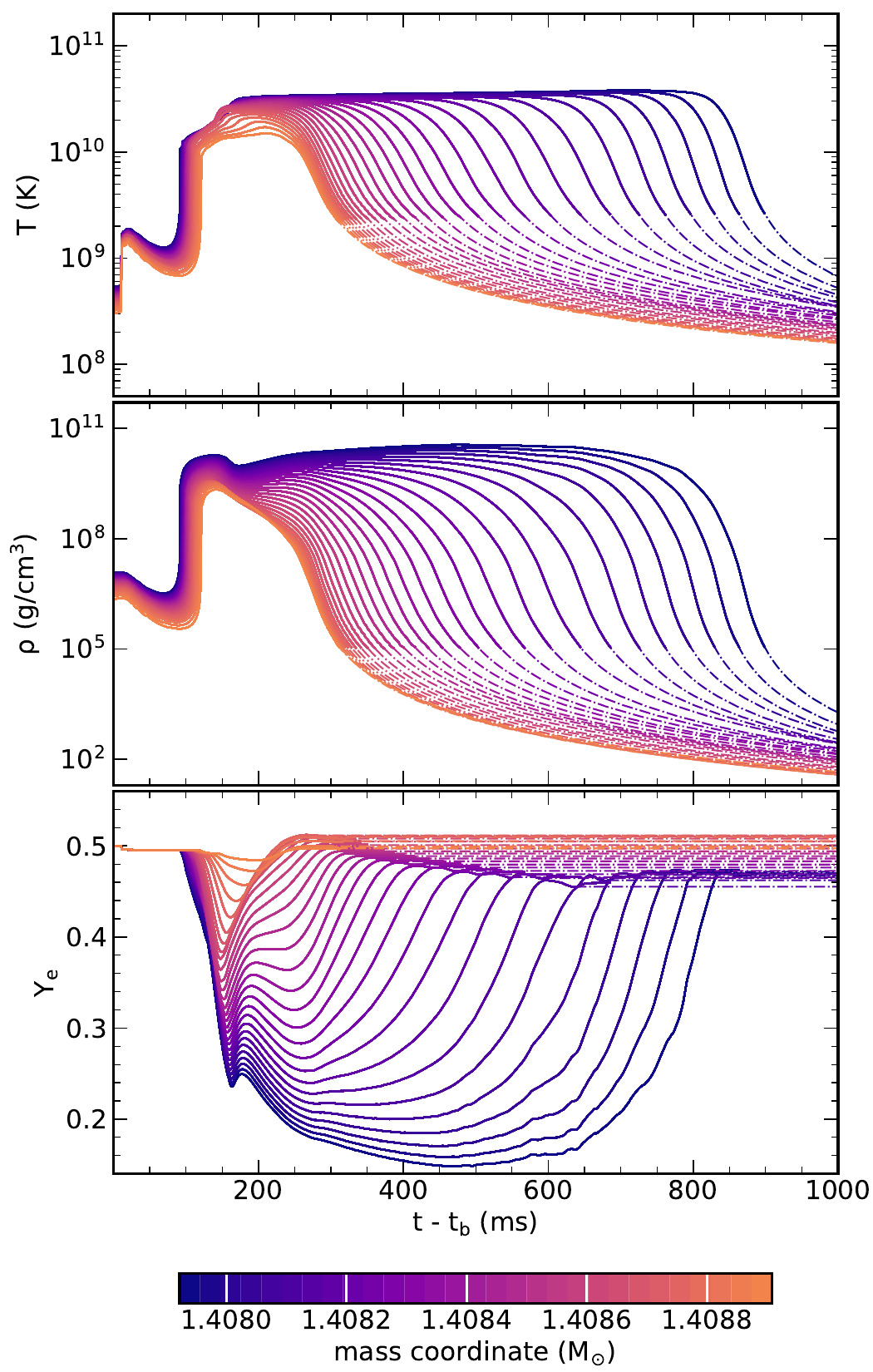}}
    \subfigure[SFHx model]{\includegraphics[width=0.45\textwidth]{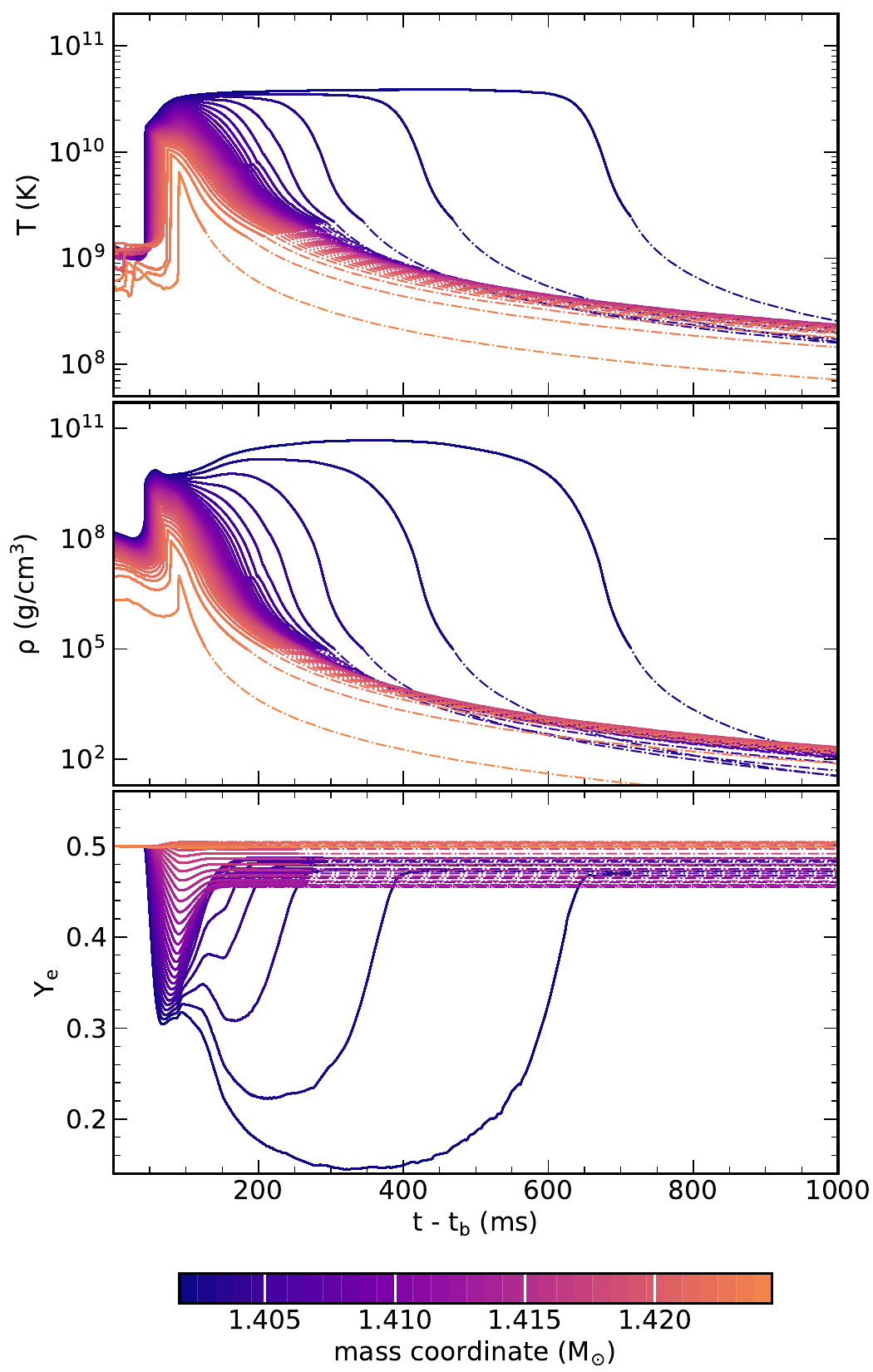}}
    \caption{Time evolution of temperature $T$ (upper panel), density $\rho$ (middle panel), and electron fraction $Y_e$ (lower panel) of the ejected trajectories, for two EoS models: a) LNS, b) SFHx. The trajectories are plotted in dashed-dotted lines after extrapolation starts.}
    \label{fig:ejecta trajectories}
\end{figure*}

The envelope of the WD is expelled after the accretion shock breaks out of the star's surface. After that, a neutrino-driven wind is developed to eject the material from the surface of the star continuously. The time evolution of ejecta mass after the core bounce in different models is displayed in Figure~\ref{fig:ejecta mass evolution}. The mass outflow properties, especially the ejecta mass, are sensitive to the EoS chosen in different models. The ejecta mass is generally larger when an EoS with higher stiffness is adopted. At the end of hydrodynamic simulations, the ejecta mass of AIC is $\sim10^{-3} (10^{-2})${\Ms} for the models using low (intermediate and high) stiffness EoSs. We map the temperature, density, and electron fraction of the ejecta to about $30$ trajectories with a conserved enclosed mass throughout the simulations. The ejected trajectories from the LNS and SFHx models are displayed in Figure~\ref{fig:ejecta trajectories} to illustrate the EoS stiffness dependence of the ejecta properties.

\begin{figure*}[t!]
    \centering
    \subfigure[LNS model] {\includegraphics[width=0.45\textwidth]{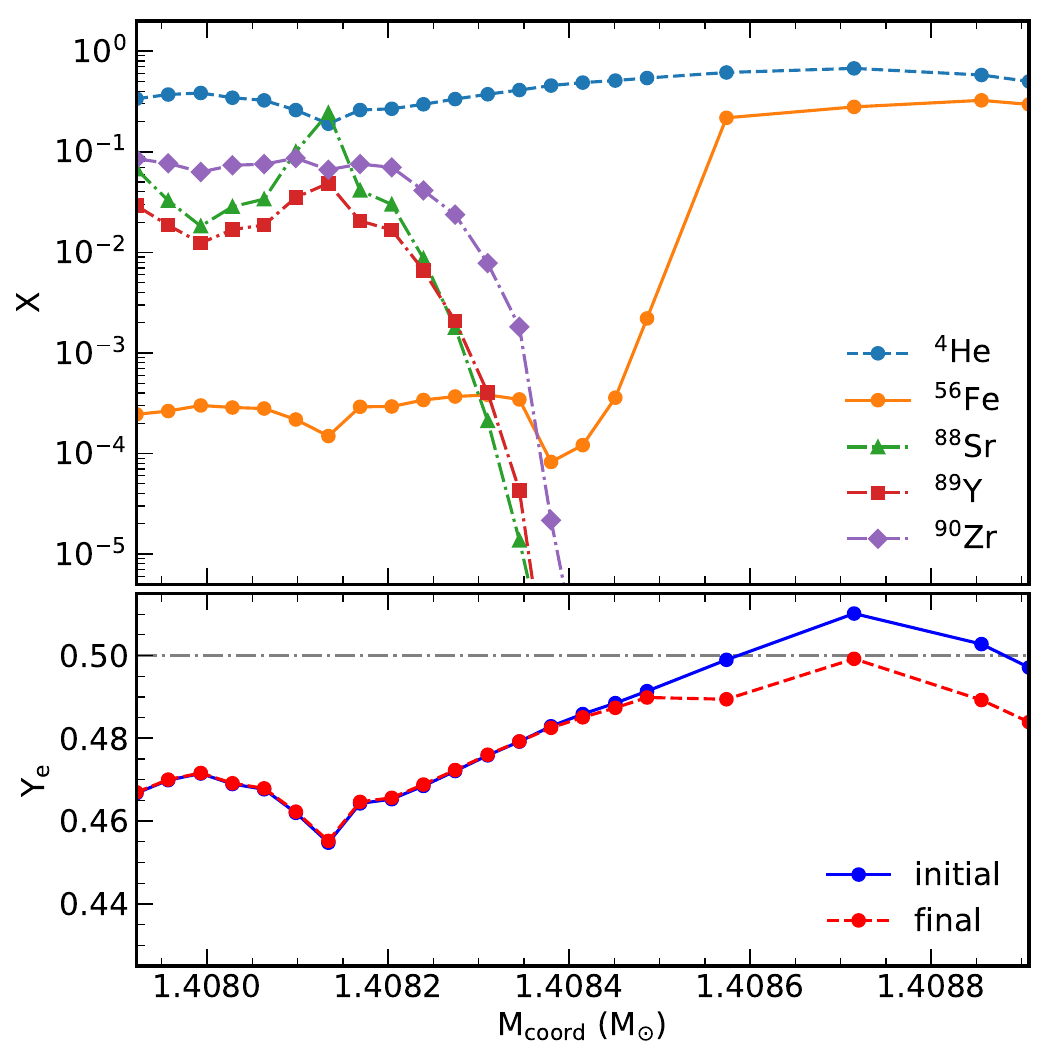}}
    \subfigure[SFHx model]{\includegraphics[width=0.45\textwidth]{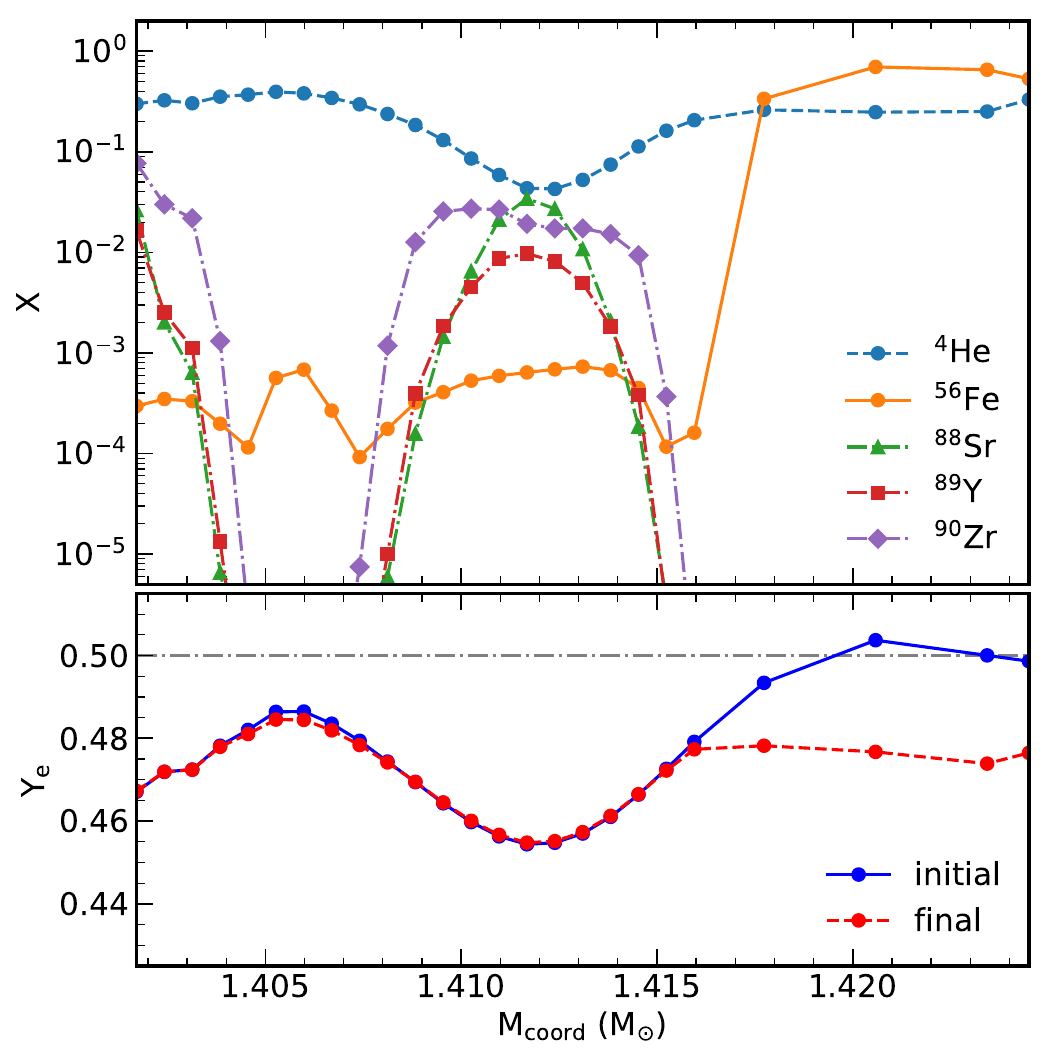}\label{fig:isotope and Ye (SFHx)}}
    \caption{Upper panels: mass fractions $X$ of a few characteristic isotopes as the end products of nucleosynthesis. Lower panels: electron fraction $Y_e$ at the beginning (blue) and at the end (red) of nucleosynthesis calculations. The above curves are plotted vs. the mass coordinate $M_{\text{coord}}$ of the ejected trajectories, for two EoS models: a) LNS, b) SFHx. The grey dash dotted horizontal lines in the lower panels indicate $Y_e=0.5$ with zero neutron excess.}
    \label{fig:isotope and Ye}
\end{figure*}

We start the nucleosynthesis calculations when the temperature of an ejected trajectory drops below $10${\GK}, and the NSE is assumed to determine the initial mass fraction of each isotope. The initial compositions are predominantly neutrons, protons, and alpha particles. The initial neutron excess is then quantified by the electron fraction $Y_e$. For the SFHo, SFHx, and STOS models, the outermost few ejected trajectories never exceed $10${\GK} throughout the simulations (see the upper panel in Figure~\ref{fig:isotope and Ye (SFHx)} for the SFHx model as an example). We start the nucleosynthesis calculations for these few trajectories after the highest temperatures ($\gtrsim5${\GK}) are reached.

\begin{figure*}[!b]
    \centering
    \subfigure[LNS model]  {\includegraphics[width=0.95\textwidth]{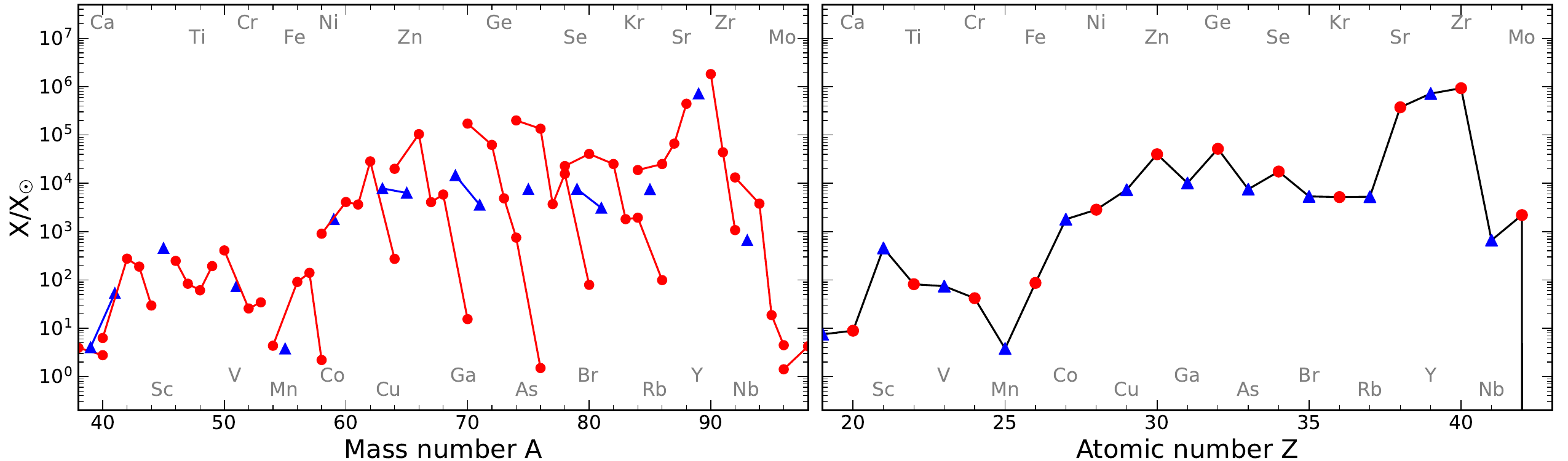}}
    \subfigure[NRAPR model]{\includegraphics[width=0.95\textwidth]{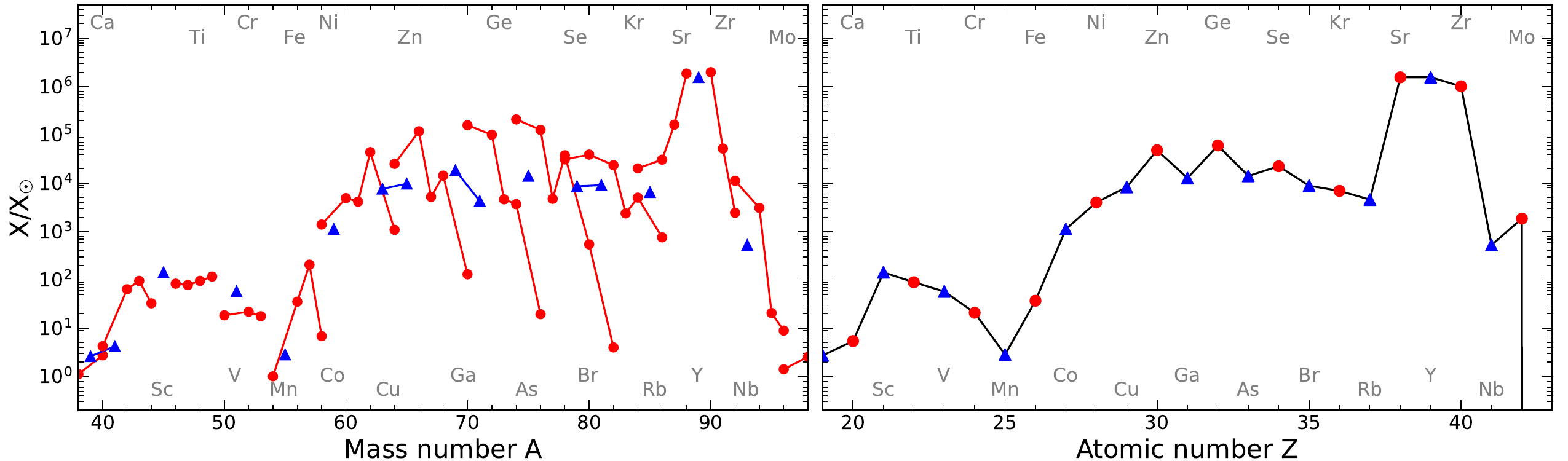}}
    \subfigure[SFHo model] {\includegraphics[width=0.95\textwidth]{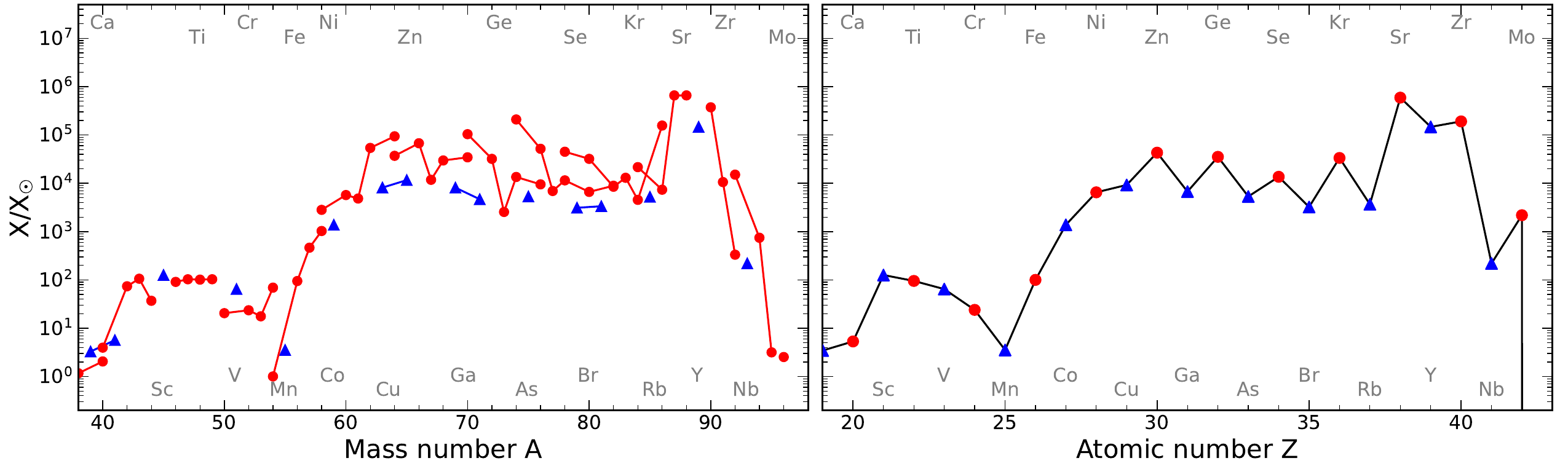}}
    \subfigure[SFHx model] {\includegraphics[width=0.95\textwidth]{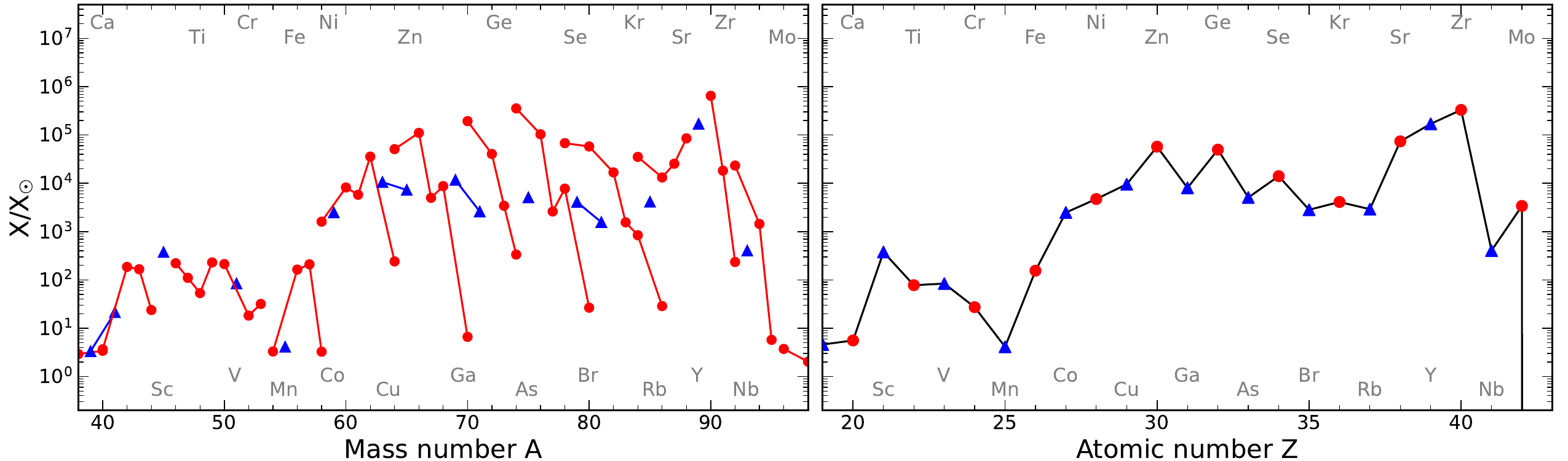}}
\end{figure*}

\begin{figure*}[t!]
    \centering
    \subfigure[STOS model] {\includegraphics[width=0.95\textwidth]{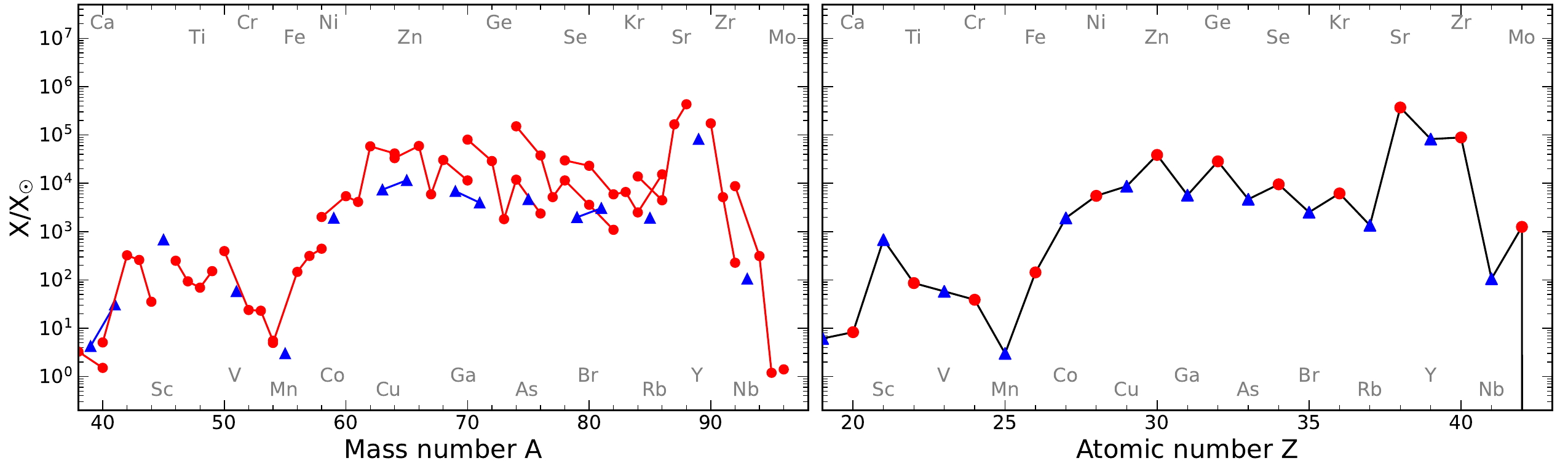}}
    \caption{Mass fractions $X$ of isotopes (left panel) and elements (right panel) over the solar abundances $X_\odot$ \citep{2019arXiv191200844L} in the end of nucleosynthesis calculations for all models. The isotopes with the same atomic number in the left panel are connected by lines. The odd-$Z$ (even-$Z$) elements are indicated by blue triangle (red circle) markers. Only the overproduced isotopes and elements ($X/X_\odot > 1$) are displaced.}
    \label{fig:overproduction of end products}
\end{figure*}

The mass fractions of $^{4}\text{He}$, $^{56}\text{Fe}$, $^{88}\text{Sr}$, $^{89}\text{Y}$, and $^{90}\text{Zr}$ versus the mass coordinate $M_{\text{coord}}$ of the ejected trajectories in the end of the nucleosynthesis calculations are displayed in the upper panel of Figure~\ref{fig:isotope and Ye} for the LNS and SFHx models. The synthesis of heavy isotopes in the AIC ejecta is mainly contributed by the $(n,\gamma)$ and $(\alpha,\gamma)$ processes. For the high initial $Y_e$ ejecta, a significant amount of $^{56}\text{Ni}$ (which decays to form $^{56}\text{Fe}$ eventually) is produced through a series of $(\alpha,\gamma)$ processes. After the freezeout of thermonuclear reactions, the proton-rich isotopes synthesized near the iron peak undergo weak decays to form stable isotopes, resulting in the drop in the final $Y_e$ (see the lower panel of Figure~\ref{fig:isotope and Ye}). For the ejected trajectories with initial $Y_e$ significantly lower than $0.5$, the yield of $^{56}\text{Ni}$ is drastically reduced. Owing to the relatively large initial neutron excess, stable neutron-rich isotopes of Ni (e.g., $^{58}\text{Ni}$ and $^{60}\text{Ni}$) are primarily produced. In addition, the first neutron-capture peak elements (Sr, Y, and Zr) are abundantly synthesized through the $r$-process in these trajectories. Unlike the high initial $Y_e$ ejecta, the resulting isotopes produced in the low initial $Y_e$ ejecta are mostly stable, and the initial $Y_e$ can be retained throughout the nucleosynthesis calculation. The overall yield of $^{56}\text{Ni}$, the major radioactive isotope that powers a supernova light curve, is $\sim10^{-4} (10^{-3})${\Ms} for the models using low (intermediate and high) stiffness EoSs. Therefore, the EM light curve of AIC is at least 2 orders dimmer than those of Type Ia SNe, with a canonical yield of $^{56}\text{Ni}$ being about $0.63${\Ms} \citep{2020ApJ...895..138K}.

To compare the production curve of AIC ejecta with the solar abundances, we plot the mass fractions $X$ of isotopes and elements relative to the solar abundances $X_\odot$ \citep{2019arXiv191200844L} for all models in Figure~\ref{fig:overproduction of end products}. The elements between the iron peak and the first neutron-capture peak are abundantly synthesized. We notice that Sr, Y, and Zr are significantly overproduced with $X/X_\odot\sim10^6$. The overproduction of even-$Z$ elements is, in general, more significant than those of odd-$Z$ elements. The yield of isotopes beyond Mo is negligible. The yield of isotopes for all models is shown in Appendix~\ref{app: yield of stable isotopes}.

\subsection{Constraint on event rate} \label{subsection: Constraint on event rate}

In this section, we follow \cite{2009ApJ...695..208W} to constrain the event rate of AICs from its chemical elements production. We firstly assume that all $^{90}\text{Zr}$ in nature are synthesized through AIC. The CCSNe from the progenitors heavier than $10${\Ms} produce $M_{\text{CCSN}} \left( ^{16}\text{O} \right)$ = $1.5${\Ms} of $^{16}\text{O}$ per event, and the production of $^{16}\text{O}$ through AIC is negligible. Let the event rate of AICs relative to that of CCSNe be $f_{\text{CCSN}}$. We have the relation
\begin{equation} \label{eq:AIC rate to CCSN}
    f_{\text{CCSN}} = \frac{X_{\odot} \left( ^{90}\text{Zr} \right) / X_{\odot} \left( ^{16}\text{O} \right)}{M_{\text{AIC}} \left( ^{90}\text{Zr} \right) / M_{\text{CCSN}} \left( ^{16}\text{O} \right)},
\end{equation}
where $X_{\odot} \left( ^{90}\text{Zr} \right) = 1.414\times10^{-8}$ and $X_{\odot} \left( ^{16}\text{O} \right) = 7.422\times10^{-3}$ are the solar abundances of $^{90}\text{Zr}$ and $^{16}\text{O}$, respectively \citep{2019arXiv191200844L}.

The representative isotope produced through Type Ia SNe, on the other hand, is $^{56}\text{Fe}$ from the weak decay of $^{56}\text{Ni}$. Analogous to Equation~(\ref{eq:AIC rate to CCSN}), the event rate of AICs relative to that of Type Ia SNe $f_{\text{SNIa}}$ is given by
\begin{equation} \label{eq:AIC rate to Type Ia SN}
    f_{\text{SNIa}} = \frac{X_{\odot} \left( ^{90}\text{Zr} \right) / X_{\odot} \left( ^{56}\text{Fe} \right)}{M_{\text{AIC}} \left( ^{90}\text{Zr} \right) / M_{\text{SNIa}} \left( ^{56}\text{Fe} \right)},
\end{equation}
where $X_{\odot} \left( ^{56}\text{Fe} \right) = 1.253\times10^{-3}$ is the solar abundance of $^{56}\text{Fe}$ \citep{2019arXiv191200844L}. We further assume that a Type Ia SN typically produces $M_{\text{SNIa}} \left( ^{56}\text{Fe} \right)$ = $0.63${\Ms} of $^{56}\text{Fe}$ \citep{2020ApJ...895..138K}, and is the primary production channel of $^{56}\text{Fe}$ among all SNe \citep{2016ApJ...825..136D}. The numerical values of $f_{\text{CCSN}}$ and $f_{\text{SNIa}}$ for different models are calculated and listed in Table~\ref{tab:model stiffness and remarkable results}. As discussed in Section~\ref{subsection: Nucleosynthesis}, the yield of $^{90}\text{Zr}$, and hence the event rate of AIC derived, are sensitive to the initial electron fraction $Y_e$ in the nucleosynthesis calculations. We show in Appendix~\ref{app: Ye test} that our results are valid up to a factor of $2$ given an uncertainty of $2\,\%$ in the initial $Y_e$ of the ejecta.

\section{Discussions} \label{section 4}

To account for the uncertainties in hot and dense nuclear matter properties, we perform the hydrodynamic simulations of AIC using five different EoSs. It is realised that the ejecta masses of AIC could differ by an order of magnitude among the EoSs presented (see Figure~\ref{fig:ejecta mass evolution}). Accordingly, the yield of heavy stable isotopes varies by an order among these models in spite of similar production curves relative to the solar abundances indicated in Figure~\ref{fig:overproduction of end products}. In general, the ejecta mass and yield of heavy stable isotopes are larger when a stiffer EoS is used. Hence, the constraints on AIC event rate relative to other supernovae derived in Section~\ref{subsection: Constraint on event rate} are also EoS dependent. From our nucleosynthesis calculations, the upper bound of AIC event rate is $\sim1-10\,\%$ relative to the Type Ia SN event rate, and the lower end of such results are consistent with estimations by previous studies \citep{2010MNRAS.409..846D,2018MNRAS.481..439W}.

We assume that AIC contributes to galactic chemical evolution, and is the major source of the first neutron-capture peak elements. As already pointed out in Section~\ref{subsection: Supernova Nucleosynthesis}, nevertheless, these elements could be abundantly synthesized through ECSNe as well. The production curves of AIC we obtained are similar to those of ECSNe presented in \cite{2009ApJ...695..208W}. Consequently, there is a degeneracy in the contribution to the solar abundances of these elements from both types of supernova. The ratios of AIC to other supernovae event rates suggested in Section~\ref{subsection: Constraint on event rate} are, therefore, conservative upper bounds of its occurrence. Further effort is needed to break the degeneracy between AIC and ECSNe.

AIC could be well distinguished from standard Type Ia SNe from the associated bright neutrino signals. With faint EM signals, AIC could be further distinguished from typical CCSNe with characteristic light curves despite similar neutrino signals. Additional efforts are required to distinguish AIC and ECSNe. For instance, radiative transfer calculations may reveal the differences in EM light curve of AIC and ECSNe with an optically thick outer envelope. After ECSN explosions, the shock heats the outer envelope, which additional powers the EM light curve of ECSNe. The compact remnants of AIC may also be distinguishable from those formed by ECSNe based on the huge difference in their progenitor masses. Meanwhile, massive highly magnetized rotating WDs are proposed as soft gamma ray repeaters and anomalous X-ray pulsars \citep{2012PASJ...64...56M}. The AIC progenitors may, therefore, be identified through EM observations before the gravitational collapse if they are highly magnetized and rapidly rotating.

The multi-dimensional effects of the explosion dynamics and ejecta properties are not investigated in this work. It is expected that WDs are rapidly rotating because of angular momentum transfer during mass accretion. A super-Chandrasekhar WD being rotationally supported may be formed as the AIC progenitor, and the explosion dynamics of such an object may be different from our models. An accretion disk formed after the AIC of a rapidly rotating progenitor \citep[see][]{2009MNRAS.396.1659M} is also neglected in our nucleosynthesis calculations. Apart from rotation, convection is also another significant multi-dimensional effect on AIC. The post-bounce convective flow may noticeably modify the electron fraction of the ejecta, and hence gives rise to distinct end products of nucleosynthesis. Moreover, the presence of a magnetic field is a potentially important factor to be considered in multi-dimensional magnetohydrodynamic simulations. The collapse of a WD may be delayed by magnetic pressure before core bounce. In addition, a relativistic jet may be launched after a PNS with a very strong magnetic field is formed. The overall production curve may be affected by the nucleosynthesis in the ejecta along the jet as suggested by \cite{2022MNRAS.514.6011B}.

The population of AIC progenitors, namely oxygen-neon-magnesium WDs, among all WDs remains unclear. Self-consistent stellar evolution simulations are required to precisely determine whether the degenerate oxygen-neon-magnesium core of a super-AGB star would collapse without leaving a WD behind. Such simulations could also reveal the chemical composition, especially Ne and Mg that are responsible for electron capture, of the WDs formed after the super-AGB stage. Furthermore, the fate of oxygen-neon-magnesium WDs evolving towards the Chandrasekhar limit through mass accretion remains unclear. For example, \cite{2015MNRAS.453.1910S} discussed the possibility of thermonuclear explosion of WDs after mass accretion. The later work by \cite{PhysRevLett.123.262701} showed that the thermonuclear runaway of oxygen is enhanced if the second-forbidden electron capture on neon and fluorine are included, and the WDs are partially disrupted by the oxygen deflagration waves. These works indicate that more accurate nuclear reaction rate calculations are required to determine whether the WDs will eventually collapse or undergo a thermonuclear explosion.

\section{Conclusions} \label{section 5}

We perform hydrodynamic simulations of AIC coupled with a neutrino transport scheme using various EoSs with different stiffness, and we investigate the yield of chemical elements in the supernova ejecta by post-processing nucleosynthesis calculations. We find that the ejecta mass and yield of heavy elements are sensitive to the stiffness of EoSs, while the trend of production curves is universal. The first neutron-capture peak elements could be overproduced by $\sim10^{6}$ relative to the solar abundances. From the yield of $^{90}\text{Zr}$, we infer that the AIC event rate cannot exceed $\sim10\,\%$ of the CCSNe and Type Ia SNe event rates to be consistent with solar abundances observation.

\section{Acknowledgement}

We thank Ken’ichi Nomoto and Shuai Zha for the helpful discussions. We acknowledge F. X. Timmes for making the Torch nuclear reaction network calculation code open-source and the explanation about its usage. We thank P. Möller for providing the publicly available isotope weak decay rates table.

This work is partially supported by grants from the Research Grant Council of the Hong Kong Special Administrative Region, China (Project Nos. 14300320 and 14304322). This material is based upon work supported by the National Science Foundation under Grant AST-2316807.

\appendix

\section{Yield of stable isotopes} \label{app: yield of stable isotopes}

The yield of stable isotopes of the LNS, NRAPR, SFHo, SFHx, and STOS models are tabulated below.

\begin{table*}
    \label{tab:yield of stable isotopes}
    \caption{Yield of stable isotopes in units of {\Ms}.}
\begin{minipage}{0.5\textwidth}
\centering
\begin{tabular}{c|ccccc}
    \hline \hline
  isotope & LNS      & NRAPR    & SFHo     & SFHx     & STOS     \\
    \hline
    $^{1}\text{H}$ & 4.19e-06 & 4.96e-10 & 1.59e-11 & 2.24e-05 & 1.03e-04 \\ 
    $^{4}\text{He}$ & 4.78e-04 & 3.17e-04 & 5.18e-03 & 5.35e-03 & 5.75e-03 \\
    $^{12}\text{C}$ & 6.52e-08 & 6.50e-08 & 1.75e-06 & 1.40e-06 & 9.33e-07 \\ 
    $^{16}\text{O}$ & 7.55e-09 & 2.26e-09 & 8.38e-09 & 5.65e-08 & 1.26e-07 \\ 
    $^{17}\text{O}$ & 2.77e-10 & 3.46e-14 & 6.85e-17 & 1.74e-09 & 7.21e-09 \\ 
    $^{18}\text{O}$ & 1.72e-11 & 3.26e-18 & 0.00e+00 & 1.04e-10 & 7.82e-10 \\ 
    $^{20}\text{Ne}$ & 1.22e-09 & 1.77e-09 & 1.13e-08 & 1.74e-08 & 2.16e-08 \\ 
    $^{21}\text{Ne}$ & 1.99e-11 & 8.37e-12 & 4.34e-14 & 1.14e-10 & 5.11e-10 \\ 
    $^{22}\text{Ne}$ & 5.34e-09 & 1.93e-12 & 1.03e-17 & 3.51e-08 & 9.26e-08 \\ 
    $^{23}\text{Na}$ & 1.57e-10 & 2.79e-14 & 3.66e-13 & 1.50e-09 & 3.35e-09 \\ 
    $^{24}\text{Mg}$ & 6.69e-10 & 1.32e-09 & 1.65e-08 & 1.57e-08 & 8.36e-09 \\ 
    $^{25}\text{Mg}$ & 2.76e-10 & 5.92e-10 & 6.00e-12 & 7.00e-10 & 3.72e-09 \\ 
    $^{26}\text{Mg}$ & 1.09e-08 & 8.47e-11 & 1.14e-11 & 7.69e-08 & 1.63e-07 \\ 
    $^{27}\text{Al}$ & 6.86e-09 & 6.70e-11 & 1.26e-09 & 4.78e-08 & 1.51e-07 \\ 
    $^{28}\text{Si}$ & 6.43e-09 & 1.06e-08 & 1.29e-07 & 1.09e-07 & 9.83e-08 \\ 
    $^{29}\text{Si}$ & 1.76e-09 & 4.28e-10 & 3.32e-10 & 8.08e-09 & 4.86e-08 \\ 
    $^{30}\text{Si}$ & 4.78e-08 & 9.22e-10 & 2.30e-08 & 3.82e-07 & 8.62e-07 \\ 
    $^{31}\text{P}$ & 3.27e-09 & 9.54e-10 & 2.78e-08 & 4.63e-08 & 6.49e-08 \\ 
    $^{32}\text{S}$ & 1.25e-08 & 2.38e-08 & 3.82e-07 & 2.57e-07 & 1.85e-07 \\ 
    $^{33}\text{S}$ & 4.15e-09 & 2.03e-09 & 6.96e-09 & 1.60e-08 & 5.46e-08 \\ 
    $^{34}\text{S}$ & 8.55e-08 & 5.73e-09 & 1.72e-07 & 9.17e-07 & 1.68e-06 \\ 
    $^{35}\text{Cl}$ & 1.01e-08 & 4.27e-09 & 1.54e-07 & 1.64e-07 & 2.16e-07 \\ 
    $^{37}\text{Cl}$ & 3.13e-09 & 5.94e-10 & 1.55e-08 & 5.07e-08 & 9.30e-08 \\ 
    $^{36}\text{Ar}$ & 1.51e-08 & 2.29e-08 & 4.20e-07 & 2.46e-07 & 3.26e-07 \\ 
    $^{38}\text{Ar}$ & 6.49e-08 & 1.65e-08 & 5.67e-07 & 1.11e-06 & 1.39e-06 \\ 
    $^{40}\text{Ar}$ & 1.25e-11 & 1.12e-11 & 2.74e-10 & 3.58e-10 & 1.76e-10 \\ 
    $^{39}\text{K}$ & 1.49e-08 & 8.72e-09 & 3.60e-07 & 2.87e-07 & 4.06e-07 \\ 
    $^{41}\text{K}$ & 1.51e-08 & 1.07e-09 & 4.72e-08 & 1.40e-07 & 2.24e-07 \\ 
    $^{40}\text{Ca}$ & 3.95e-07 & 2.40e-07 & 7.34e-06 & 5.34e-06 & 8.21e-06 \\ 
    $^{42}\text{Ca}$ & 1.21e-07 & 2.53e-08 & 9.59e-07 & 1.90e-06 & 3.66e-06 \\ 
    $^{43}\text{Ca}$ & 1.77e-08 & 8.04e-09 & 2.93e-07 & 3.62e-07 & 6.24e-07 \\ 
    $^{44}\text{Ca}$ & 4.40e-08 & 4.37e-08 & 1.62e-06 & 8.21e-07 & 1.35e-06 \\ 
    $^{45}\text{Sc}$ & 1.96e-08 & 5.52e-09 & 1.60e-07 & 3.75e-07 & 7.52e-07 \\ 
    $^{46}\text{Ti}$ & 6.56e-08 & 1.98e-08 & 7.10e-07 & 1.36e-06 & 1.69e-06 \\ 
    $^{47}\text{Ti}$ & 2.03e-08 & 1.71e-08 & 7.40e-07 & 6.24e-07 & 5.85e-07 \\ 
    $^{48}\text{Ti}$ & 1.50e-07 & 2.11e-07 & 7.36e-06 & 3.04e-06 & 4.37e-06 \\ 
    $^{49}\text{Ti}$ & 3.57e-08 & 1.95e-08 & 5.62e-07 & 9.85e-07 & 7.21e-07 \\ 
    $^{50}\text{Ti}$ & 2.68e-12 & 4.38e-12 & 2.06e-09 & 1.36e-11 & 2.59e-11 \\ 
    $^{51}\text{V}$ & 2.95e-08 & 2.06e-08 & 7.57e-07 & 7.73e-07 & 5.91e-07 \\ 
    $^{50}\text{Cr}$ & 3.30e-07 & 1.34e-08 & 4.89e-07 & 3.99e-06 & 8.21e-06 \\ 
    $^{52}\text{Cr}$ & 4.16e-07 & 3.19e-07 & 1.13e-05 & 6.86e-06 & 9.92e-06 \\ 
    $^{53}\text{Cr}$ & 6.45e-08 & 2.98e-08 & 9.76e-07 & 1.38e-06 & 1.11e-06 \\ 
    $^{54}\text{Cr}$ & 1.49e-10 & 1.46e-10 & 9.67e-07 & 9.91e-10 & 6.70e-08 \\ 
    $^{55}\text{Mn}$ & 5.37e-08 & 3.61e-08 & 1.48e-06 & 1.35e-06 & 1.09e-06 \\ 
    $^{54}\text{Fe}$ & 3.39e-07 & 7.05e-08 & 2.31e-06 & 5.97e-06 & 9.94e-06 \\ 
    $^{56}\text{Fe}$ & 1.15e-04 & 4.03e-05 & 3.55e-03 & 4.82e-03 & 4.79e-03 \\ 
    $^{57}\text{Fe}$ & 4.20e-06 & 5.54e-06 & 4.09e-04 & 1.46e-04 & 2.41e-04 \\ 
    $^{58}\text{Fe}$ & 8.89e-09 & 2.49e-08 & 1.23e-04 & 3.06e-07 & 4.59e-05 \\  
    \hline
\end{tabular}
\end{minipage}
\begin{minipage}{0.5\textwidth}
\centering
\begin{tabular}[t]{c|ccccc}
    \hline \hline
  isotope & LNS      & NRAPR    & SFHo     & SFHx     & STOS     \\
    \hline 
    $^{59}\text{Co}$ & 6.81e-06 & 3.83e-06 & 1.54e-04 & 2.17e-04 & 1.86e-04 \\
    $^{58}\text{Ni}$ & 4.93e-05 & 6.83e-05 & 4.52e-03 & 2.03e-03 & 2.80e-03 \\ 
    $^{60}\text{Ni}$ & 8.92e-05 & 9.64e-05 & 3.66e-03 & 4.13e-03 & 3.02e-03 \\ 
    $^{61}\text{Ni}$ & 3.49e-06 & 3.59e-06 & 1.37e-04 & 1.29e-04 & 1.01e-04 \\ 
    $^{62}\text{Ni}$ & 8.85e-05 & 1.23e-04 & 4.95e-03 & 2.57e-03 & 4.65e-03 \\ 
    $^{64}\text{Ni}$ & 2.23e-07 & 7.97e-07 & 2.26e-03 & 4.56e-06 & 8.70e-04 \\ 
    $^{63}\text{Cu}$ & 5.18e-06 & 4.56e-06 & 1.58e-04 & 1.62e-04 & 1.25e-04 \\ 
    $^{65}\text{Cu}$ & 1.92e-06 & 2.66e-06 & 1.04e-04 & 5.11e-05 & 9.02e-05 \\ 
    $^{64}\text{Zn}$ & 2.25e-05 & 2.55e-05 & 1.23e-03 & 1.33e-03 & 9.58e-04 \\ 
    $^{66}\text{Zn}$ & 6.81e-05 & 6.98e-05 & 1.30e-03 & 1.67e-03 & 9.94e-04 \\ 
    $^{67}\text{Zn}$ & 3.95e-07 & 4.55e-07 & 3.37e-05 & 1.12e-05 & 1.47e-05 \\ 
    $^{68}\text{Zn}$ & 2.63e-06 & 5.80e-06 & 3.93e-04 & 9.10e-05 & 3.49e-04 \\ 
    $^{70}\text{Zn}$ & 2.45e-10 & 1.86e-09 & 1.61e-05 & 2.44e-09 & 4.67e-06 \\ 
    $^{69}\text{Ga}$ & 6.20e-07 & 7.08e-07 & 1.02e-05 & 1.15e-05 & 7.46e-06 \\ 
    $^{71}\text{Ga}$ & 1.04e-07 & 1.12e-07 & 4.00e-06 & 1.75e-06 & 2.95e-06 \\ 
    $^{70}\text{Ge}$ & 8.45e-06 & 6.98e-06 & 1.50e-04 & 2.20e-04 & 1.01e-04 \\ 
    $^{72}\text{Ge}$ & 4.20e-06 & 6.12e-06 & 6.36e-05 & 6.31e-05 & 5.00e-05 \\ 
    $^{73}\text{Ge}$ & 9.40e-08 & 8.04e-08 & 1.45e-06 & 1.53e-06 & 8.95e-07 \\ 
    $^{74}\text{Ge}$ & 6.92e-08 & 3.06e-07 & 3.67e-05 & 7.15e-07 & 2.81e-05 \\ 
    $^{76}\text{Ge}$ & 3.00e-11 & 3.51e-10 & 5.62e-06 & 1.65e-10 & 1.22e-06 \\ 
    $^{75}\text{As}$ & 9.73e-08 & 1.64e-07 & 2.03e-06 & 1.52e-06 & 1.55e-06 \\ 
    $^{74}\text{Se}$ & 2.44e-07 & 2.30e-07 & 7.51e-06 & 1.00e-05 & 4.73e-06 \\ 
    $^{76}\text{Se}$ & 1.81e-06 & 1.53e-06 & 2.05e-05 & 3.22e-05 & 1.30e-05 \\ 
    $^{77}\text{Se}$ & 4.15e-08 & 4.79e-08 & 2.29e-06 & 6.77e-07 & 1.49e-06 \\ 
    $^{78}\text{Se}$ & 5.54e-07 & 1.21e-06 & 1.20e-05 & 6.36e-06 & 1.04e-05 \\ 
    $^{80}\text{Se}$ & 6.03e-09 & 3.72e-08 & 1.49e-05 & 4.70e-08 & 7.03e-06 \\ 
    $^{82}\text{Se}$ & 3.13e-12 & 4.97e-11 & 3.63e-06 & 1.23e-11 & 3.87e-07 \\ 
    $^{79}\text{Br}$ & 1.07e-07 & 1.09e-07 & 1.29e-06 & 1.33e-06 & 7.12e-07 \\ 
    $^{81}\text{Br}$ & 4.33e-08 & 1.15e-07 & 1.38e-06 & 5.08e-07 & 1.10e-06 \\ 
    $^{78}\text{Kr}$ & 9.57e-09 & 1.18e-08 & 5.57e-07 & 6.59e-07 & 3.21e-07 \\ 
    $^{80}\text{Kr}$ & 1.10e-07 & 9.59e-08 & 2.57e-06 & 3.64e-06 & 1.61e-06 \\ 
    $^{82}\text{Kr}$ & 3.49e-07 & 2.96e-07 & 3.48e-06 & 5.41e-06 & 2.11e-06 \\ 
    $^{83}\text{Kr}$ & 2.53e-08 & 2.99e-08 & 5.36e-06 & 5.04e-07 & 2.37e-06 \\ 
    $^{84}\text{Kr}$ & 1.35e-07 & 3.15e-07 & 9.31e-06 & 1.36e-06 & 4.45e-06 \\ 
    $^{86}\text{Kr}$ & 2.12e-09 & 1.46e-08 & 9.90e-05 & 1.43e-08 & 8.44e-06 \\ 
    $^{85}\text{Rb}$ & 9.16e-08 & 7.19e-08 & 1.89e-06 & 1.17e-06 & 6.04e-07 \\ 
    $^{84}\text{Sr}$ & 5.83e-09 & 5.65e-09 & 1.97e-07 & 2.52e-07 & 1.10e-07 \\ 
    $^{86}\text{Sr}$ & 1.40e-07 & 1.55e-07 & 1.22e-06 & 1.72e-06 & 6.42e-07 \\ 
    $^{87}\text{Sr}$ & 2.64e-07 & 5.81e-07 & 7.72e-05 & 2.34e-06 & 1.69e-05 \\ 
    $^{88}\text{Sr}$ & 2.12e-05 & 8.03e-05 & 9.32e-04 & 9.45e-05 & 5.32e-04 \\ 
    $^{89}\text{Y}$ & 7.91e-06 & 1.54e-05 & 4.76e-05 & 4.32e-05 & 2.32e-05 \\ 
    $^{90}\text{Zr}$ & 2.62e-05 & 2.57e-05 & 1.58e-04 & 2.15e-04 & 6.41e-05 \\ 
    $^{91}\text{Zr}$ & 1.39e-07 & 1.49e-07 & 9.88e-07 & 1.34e-06 & 4.21e-07 \\ 
    $^{92}\text{Zr}$ & 5.26e-09 & 1.07e-08 & 4.79e-08 & 2.66e-08 & 2.84e-08 \\ 
    $^{93}\text{Nb}$ & 1.37e-09 & 9.73e-10 & 1.33e-08 & 1.92e-08 & 5.59e-09 \\ 
    $^{92}\text{Mo}$ & 1.31e-08 & 1.00e-08 & 4.41e-07 & 5.34e-07 & 2.22e-07 \\ 
    $^{94}\text{Mo}$ & 2.40e-09 & 1.76e-09 & 1.39e-08 & 2.12e-08 & 5.08e-09 \\ 
    $^{95}\text{Mo}$ & 2.07e-11 & 2.05e-11 & 1.03e-10 & 1.48e-10 & 3.40e-11 \\  
    \hline
\end{tabular}
\end{minipage}
\end{table*}

\section{Sensitivity to variation in electron fraction} \label{app: Ye test}
The production curve of AIC, especially the first neutron-capture peak elements, is sensitive to the initial electron fraction $Y_e$ of the ejecta, as the yield of neutron-rich isotopes is directly related to the initial neutron excess. In realistic supernova simulations, the $Y_e$ of ejecta may be subject to an uncertainty up to a few percents due to both finite numerical resolution and other physical effects, such as convective mixing and uncertainties of neutrino-matter interaction rates \citep[see the discussions in][for more details]{2009ApJ...695..208W}. Here, we study the sensitivity of nucleosynthesis yield to the initial $Y_e$ of the ejecta to quantify the robustness of our results.

\begin{figure*}[h!]
    \centering
    \subfigure[LNS model]  {\includegraphics[width=0.95\textwidth]{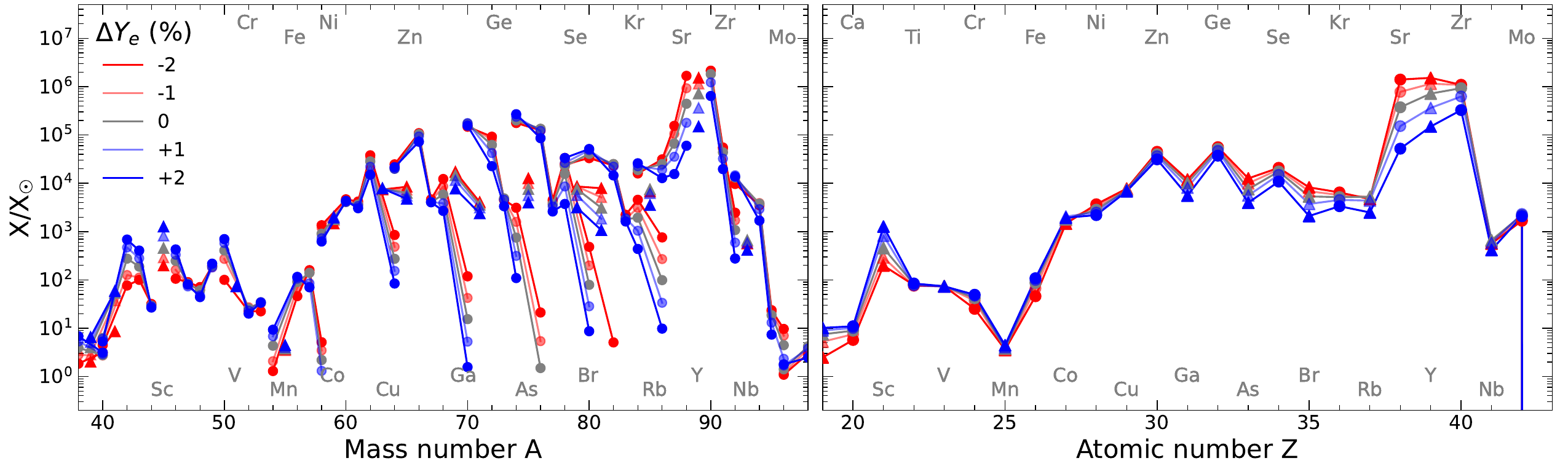}}
    \subfigure[SFHx model] {\includegraphics[width=0.95\textwidth]{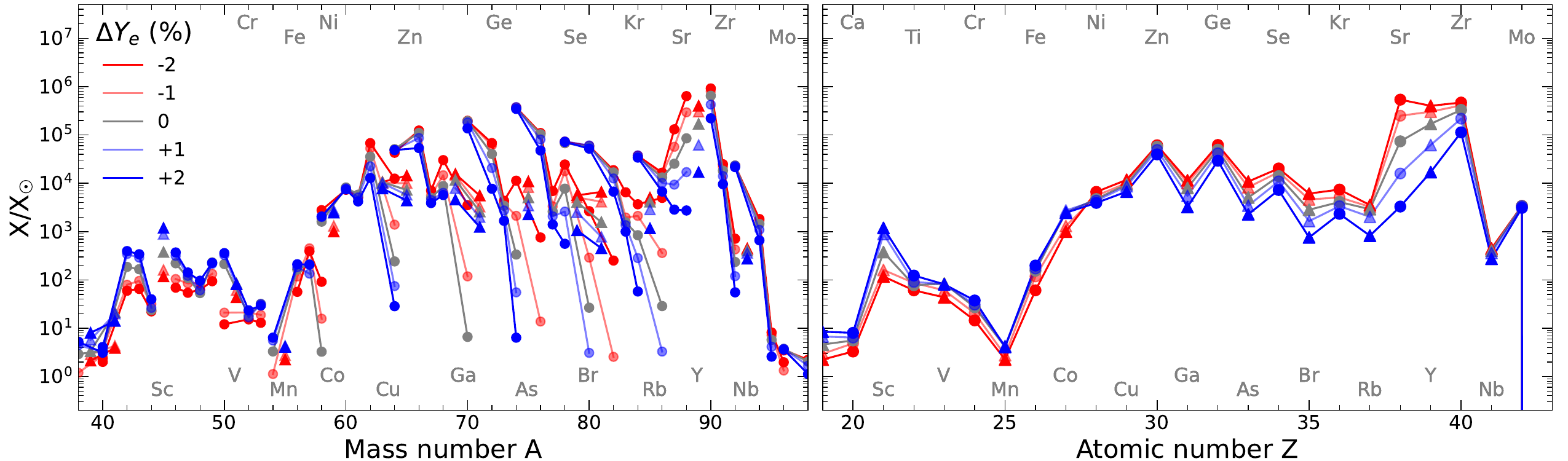}}
    \caption{Same figure as Figure~\ref{fig:overproduction of end products}, but the production curves with initial electron fraction $Y_e$ of ejecta artificially modified by $+1\,\%$ (pale blue), $+2\,\%$ (blue), $-1\,\%$ (pale red), and $-2\,\%$ (red) are attached to compare with the canonical production curve (grey) for two EoS models: a) LNS, b) SFHx.}
    \label{fig:Ye test}
\end{figure*}

We artificially modify the initial $Y_e$ of the ejecta by $\pm1\,\%$ and $\pm2\,\%$ at the beginning of nucleosynthesis calculations done in Section~\ref{subsection: Nucleosynthesis} for all models. We find that the yield of heavy elements, especially the first neutron-capture peak elements, is significantly altered. In general, the yield of these heavy elements increases (decreases) when the initial $Y_e$ decreases (increases). The comparison among the production curves with different initial $Y_e$ is shown in Figure~\ref{fig:Ye test} for the LNS and SFHx models. A wide range of initial $Y_e$, down to $Y_e\lessapprox0.46$, is covered by the AIC ejecta in all models. We notice that the yield of $^{90}\text{Zr}$ peaks at $Y_e\approx0.46$ in most of the trajectories. When we shift the initial $Y_e$ of all trajectories downward, the overall enhancement of $^{90}\text{Zr}$ production is somehow compensated by the reduction of $^{90}\text{Zr}$ production in a few trajectories with $Y_e<0.46$, and vice versa. The yield of $^{88}\text{Sr}$ ($^{89}\text{Y}$), on the other hand, decreases monotonically with $Y_e$ for $Y_e\gtrapprox0.44 (0.45)$. For the range of initial $Y_e$ covered by the AIC ejecta, a downward shifting in $Y_e$ results in a large enhancement of $^{88}\text{Sr}$ and $^{89}\text{Y}$ production in general. Hence, the yield of $^{90}\text{Zr}$ is less sensitive to $Y_e$ compared to other first neutron-capture peak elements in our models. Within an uncertainty of $\pm2\,\%$ for the initial $Y_e$, we find that the yield of $^{90}\text{Zr}$ may vary by a factor of $2$ in all models. The event rates of AIC reported in Table~\ref{tab:model stiffness and remarkable results} are, correspondingly, valid up to a factor of $2$, assuming such an uncertainty of $Y_e$.

\bibliography{sample631}{}
\bibliographystyle{aasjournal}

\end{document}